\documentclass[12pt,epsf]{article}
  \usepackage{epsfig}
  \newlength{\abstractwidth}
  \renewcommand{\thefootnote}{\fnsymbol{footnote}}
  \renewcommand{\thanks}[1]{\footnote{#1}} 
  \newcommand{\starttext}{
  \setcounter{footnote}{0}
  \renewcommand{\thefootnote}{\arabic{footnote}}}
  \renewcommand{\theequation}{\thesection.\arabic{equation}}
  \newcommand{\be}{\begin{equation}}
  \newcommand{\bea}{\begin{eqnarray}}
  \newcommand{\eea}{\end{eqnarray}}
  \newcommand{\beq}{\begin{equation}}
  \newcommand{\ee}{\end{equation}}
  \newcommand{\eeq}{\end{equation}}

  \def\ba{\begin{eqnarray}}
  \def\ea{\end{eqnarray}}

\def\14{{1\over4}}
  \def\12{{1 \over 2}}
  \def\eq{&=&}

  \def\ds{\partial_{\sigma}}
  \def\h3{h^{3\over 2}}
  
  \def\ds{de Sitter space}

  \def\ba{\begin{eqnarray}}
\def\ea{\end{eqnarray}}
\def\ba{\begin{eqnarray}}
  \def\ea{\end{eqnarray}}

  \def\12{{1 \over 2}}
  \def\eq{&=&}

  \def\ds{\partial_{\sigma}}
\def\h3{h^{3\over 2}}
  
  \def\ds{de Sitter space}

  \def\cc{cosmological constant}
  
  \def\simleq{\; \raise0.3ex\hbox{$<$\kern-0.75em
      \raise-1.1ex\hbox{$\sim$}}\; }
   \def\simgeq{\; \raise0.3ex\hbox{$>$\kern-0.75em
      \raise-1.1ex\hbox{$\sim$}}\; }

\def\cdl{Coleman De Luccia}

\def\ba{\bf{a}}

\def\h3{{\cal{H}}_3}
\def\e{\epsilon}
\def\eb{Einstein-Rosen bridge}

  \def\simleq{\; \raise0.3ex\hbox{$<$\kern-0.75em
      \raise-1.1ex\hbox{$\sim$}}\; }
   \def\simgeq{\; \raise0.3ex\hbox{$>$\kern-0.75em
      \raise-1.1ex\hbox{$\sim$}}\; }

\newcommand{\ltsim}{\protect\raisebox{-0.5ex}{$\:\stackrel{\textstyle <}
	{\sim}\:$}}
\newcommand{\gtsim}{\protect\raisebox{-0.5ex}{$\:\stackrel{\textstyle >}
	{\sim}\:$}}

\begin{document}
  \renewcommand{\theequation}{\thesection.\arabic{equation}}
  \begin{titlepage}

\bigskip\bigskip
\rightline{SITP 10/04}
\rightline{OIQP-10-01}

  \bigskip\bigskip\bigskip\bigskip

  \centerline{\Large \bf {On the Topological Phases of Eternal Inflation}}

  \bigskip\bigskip
  \bigskip\bigskip
  
\begin{center}
  {{\large Yasuhiro Sekino${}^{1,2}$, Stephen Shenker${}^{1}$,
and Leonard Susskind${}^{1}$ }}
  \bigskip

\bigskip
${}^1\;$   {\it Department of Physics, Stanford University\\
Stanford, CA 94305-4060, USA} 

\medskip

${}^2\;$ {\it Okayama Institute for Quantum Physics\\
1-9-1 Kyoyama, Okayama 700-0015, Japan}
\vspace{2cm}
  \end{center}
  \bigskip

  \bigskip\bigskip
  \begin{abstract}
Eternal inflation is a term that describes a number of different phenomena
which have been classified by Winitzki. According to Winitzki's
classification these phases can be  characterized by the topology of the
percolating structures in the inflating, ``white,''  region.

In this paper we discuss these phases, the transitions between them, and the
way they are seen by a ``Census Taker''; a hypothetical observer inside the
non-inflating, ``black,'' region.  We discuss three phases that we call,
``black island,'' ``tubular,'' and ``white island.''  The black island phase is
familiar,  comprised of rare Coleman De Luccia bubble nucleation events.
The Census Taker sees an essentially spherical boundary, described by the
conformal field theory of the FRW/CFT correspondence.   In the tubular phase
the Census Taker sees a complicated infinite genus structure composed of
arbitrarily long tubes.  The white island phase is even more mysterious from
the black side.  Surprisingly, when viewed from the non-inflating region
this phase resembles a closed, positively curved universe which eventually
collapses to a singularity. Nevertheless, pockets of eternal inflation
continue forever.

In addition there is an ``aborted'' phase in which no eternal inflation takes
place.

 Rigorous results of Chayes, Chayes, Grannan and Swindle establish the
existence of all of these phases, separated by first order transitions, in
Mandelbrot percolation, a simple model of eternal inflation.
  \end{abstract}

  \end{titlepage}
  \starttext \baselineskip=17.63pt \setcounter{footnote}{0}

\setcounter{equation}{0}
\section{Introduction}

An existence of a large landscape of vacua  strongly suggests that the early universe underwent a process of eternal inflation, during which the multiverse was populated with  bubble universes of every type. This idea is the basis for a number of  applications of  anthropic selection.

The importance of understanding  the principles and mechanisms that underlie eternal inflation, and the need to sort out, classify, and distinguish  the various phenomena that are grouped under the single heading of eternal inflation, is obvious. In fact eternal inflation is not a single phenomenon: several  types of eternal inflation---some of them very surprising---are all possible. Moreover, as the parameters of the theory are varied,  phase transitions from one to another type of eternal inflation take place\footnote{To avoid confusion, when we say that the parameters vary, we do not mean that they vary with time. The parameters are frozen for a particular model, but we are free to vary them and see how cosmic evolution changes. The term phase and phase transition does not refer to the instantaneous behavior the vacuum but to entire cosmological evolution of the multiverse, or more exactly, its asymptotically late behavior.}. Although the phases are enormously different, at the present time we cannot say which type of eternal inflation our universe evolved out of.

In the simplified version of the landscape that we will discuss, there
are only two vacua, a  metastable ancestor vacuum, which we call
the ``white'' vacuum, and a non-inflating
offspring vacuum with vanishing \cc, which we call the ``black''
vacuum. The ancestor vacuum decays by
producing bubbles of offspring vacuum~\cite{Coleman:1980aw}.
The relevant control parameter---the variable that one scans to go from one phase to another---is the   rate for  the ancestor  to decay into the offspring.  Winitzki ~\cite{Winitzki} has given a useful topological characterization of the phases based on the nature  of the percolating structures present in the inflating, ``white" region.   Using this characterization in three spatial dimensions we discuss the following phases:

\begin{itemize}

\item When the decay rate is very small the behavior is familiar. Isolated
clusters of bubbles of black offspring vacuum form and grow, but the accelerated expansion of the inflating ancestor prevents the clusters from merging. This phase of isolated offspring bubbles will be called the ``black island" phase.  Sheets of white percolate.

\item As the decay rate increases, a transition to a ``tubular'' phase occurs;
the bubbles run into each other, and percolate.  This is the original
percolation transition discussed by Guth and
Weinberg~\cite{weinberg}. But the percolation does not shut down the
eternal inflation. What does happen is that the decay products merge
into a single black tubular structure, but the tubular decay product
does not fill space. The un-decayed white region is also a percolating
tubular network that continues to inflate.
A time-like observer in the zero cosmological constant black region,
which we call the ``Census Taker,'' will eventually see the whole
tubular structure with an infinite genus boundary.

\item  An even more  bizarre phenomenon occurs as the decay rate increases further. In the ``white island'' phase the white region does not percolate. Instead it forms a collection of isolated islands surrounded by black. However, each white island is unstable to cracking---a phenomenon by which the islands eternally inflate split, inflate, and split. They never grow much larger than the Hubble scale of the inflating vacuum. Meanwhile, in the black regions, singularities form in the cracks. These singularities may merge and cause the space to crunch so that every time-like trajectory ends at a singularity.
\end{itemize}

\subsection{Preliminary Comments About Inflation}

\begin{figure}\begin{center}\includegraphics{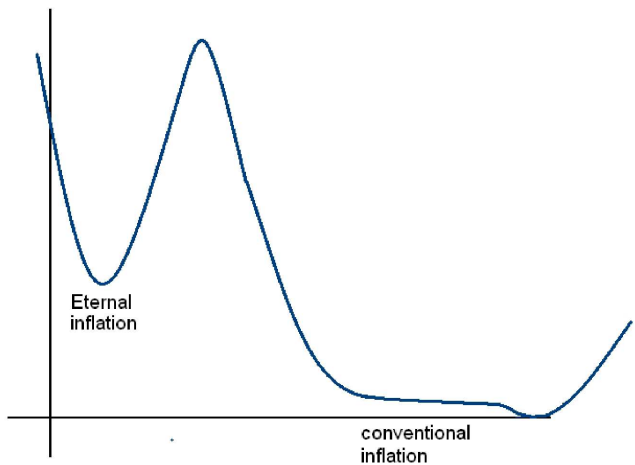}\caption{A potential that supports eternal inflation and conventional inflation.}\label{potential1}
\end{center}\end{figure}

Eternal inflation is a different phenomenon from  conventional inflation which flattened the universe and created the primordial fluctuation spectrum.
The difference  has sometimes been obscured by attempts to derive both
from a single smooth potential.  The difference  can be most easily seen
when eternal inflation occurs in a long-lived,
metastable vacuum, which decays by  quantum tunneling. A tunneling event may lead to a bubble universe which subsequently undergoes  conventional non-eternal inflation. In this case the two kinds of inflation are clearly distinct from one another.
A simple model of such behavior is obtained using a scalar potential with the shape indicated in Figure \ref{potential1}.

For the questions addressed in this paper conventional inflation plays no important role. Thus it is unnecessary to distinguish non-eternal inflation from \it no inflation at all.\rm  \   For present purposes the shallow plateau in Figure \ref{potential1} can be ignored.

Another property of the real world which will not play a role in our discussion is the tiny cosmological constant that appears to dominate today's energy density. Compared to the potential energy during eternal inflation, the cosmological constant must be more than 100 orders of magnitude smaller. It is obvious that for many purposes it may be
ignored. Moreover, there are theoretical reasons \cite{FSSY, foam, hat}
to study the behavior of transitions to terminal vacua with vanishing \cc.

In this paper we will consider the various phases, and phase  transitions, that  may occur as the parameters of the inflaton potential are varied. That at least one  transition occurs is well known, namely the transition to eternal inflation from no-eternal-inflation \cite{leonardo}.  But as we will see, there is reason to believe that eternal inflation itself has several phases.

\begin{figure}\begin{center}\includegraphics{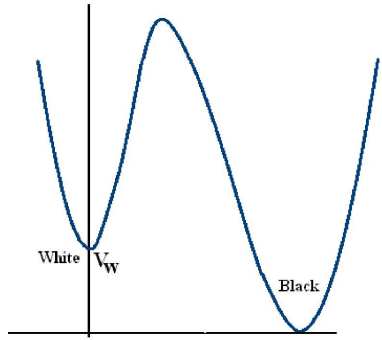}\caption{A very simple model landscape. The two minima of the  potential are called White and  Black.}\label{potential2}
\end{center}\end{figure}

For illustrative purposes we  simplify the landscape to a single scalar field, with a potential shown in Figure \ref{potential2}. The potential has two  minima, one at $\phi_w$ with positive energy $V_w$, and one at $\phi_b$ with vanishing energy. We call these vacua  $white$ and $black.$ 
Tunneling from the white vacuum to the black vacuum is possible.
In a space that is in the white vacuum, bubbles of the black vacuum
will form. The nucleation rate per unit 4-volume is defined to be $\Gamma$.
The rate for tunneling up from the zero energy black vacuum to
the positive energy white vacuum is zero.

\begin{figure}\begin{center}\includegraphics{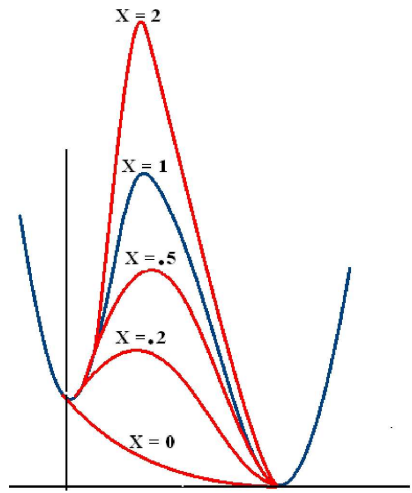}\caption{A one parameter family of potentials. For large  $X$ the white vacuum is metastable and decays by \cdl \ bubble nucleation. For small $X$ there is  only one minimum and no inflation takes place.}\label{potential3}
\end{center}\end{figure}

We will also introduce a control parameter $X$. When $X=1$ the potential
is as shown in Figure \ref{potential2}. As $X\to \infty $ the barrier
between the minima increases so that the tunneling rate tends to zero.
As $X$ decreases toward zero
the barrier between the minima decreases and eventually at $X=0$ the
potential is such that no metastable vacuum or slow-roll region
exists. (See Figure \ref{potential3}.)

A single scalar field with two minima is a vast oversimplification of the landscape. Nevertheless, it is complicated enough to see a variety of very different behaviors.  In a  multiverse based on a richer landscape the phase structure is undoubtedly far more complex than what we  describe here.

In what follows, we will assume  that the universe (or a patch of it) started everywhere white\footnote{For our purposes we don't need to know how it got there. Perhaps the simplest explanation is that it tunneled to the white minimum from some earlier vacuum with even larger energy.}, with $\phi = \phi_w$.
We are interested in classifying the various  behaviors that can occur
when the control parameter  $X$ is scanned. More precisely, in
this paper we will study the behavior as we vary the nucleation rate $\Gamma$.
In particular we would like
to know if there are discrete  transitions;  if so are they first or
second order (sudden or smooth); and finally, are there order parameters
describing  the phases. Note that by a phase we do not refer to equilibrium behavior but rather to  the properties of a cosmological  history.

Before considering the cosmological situation, let us first consider the case of an  ordinary field theory in flat space-time with the potential of Figure \ref{potential2}.
 The white vacuum is metastable and  the black vacuum is stable. Start with an initial condition such that the white vacuum fills space. Tunneling processes will take place, leading to the nucleation of bubbles of black vacuum.

In a time of order $\Gamma^{-1/4}$ space will be occupied by a collection of bubbles spaced by an average separation of $\Gamma^{-1/4}$. The bubbles will grow at the speed of light and coalesce. Thus in a time $\sim \Gamma^{-1/4}$
 the entire space will have decayed to the black configuration. It does not matter how small $\Gamma$ is; the history is always the same. The white configuration decays in a finite time to uniform black (plus some radiation).
  In the sense of this paper there is only one phase as $X$
is varied.

The situation is different in an inflating background. In that case there is a competition between the bubble nucleation
rate and the expansion rate in the inflating ancestor. For small enough
\cc \ the flat space-time picture is correct, but for larger \cc \  the
growing bubbles cannot outrun the Hubble expansion, and a transition to
eternal inflation takes place.

In terms of comoving coordinates, the radius of a bubble asymptotically
approaches a fixed size given by the horizon radius at the nucleation
time. On the future conformal infinity, bubbles are represented as
spherical regions cut out from the inflating space. Bubbles
nucleated early are represented by large spheres, and those nucleated
late by small spheres.
This gives rise to a self-similar fractal of inflating space.
This fact has been noted by Guth and Weinberg~\cite{weinberg}
in their study of the distribution of bubbles. The fractal nature
is an essential property of eternal inflation, and its importance
has been emphasized in various contexts.
See recent discussion by Garriga and Vilenkin~\cite{alex v}.

The pattern of bubbles can be described by a simple percolation model due to
Mandelbrot~\cite{Mandelbrot} who understood its essential features, which will be reviewed in Section 3.  This model was first used to study eternal inflation by Kesten (as cited in~\cite{weinberg}) and further analyzed in ~\cite{AryalVilenkin, Winitzki}.
Chayes, Chayes Durrett, Grannan and Swindle~\cite{Mandelbrot} rigorously established the existence of the phases described above characterized by their percolating structures, each separated from the other by first order transitions.

In the following sections we will review the various phases of the
Mandelbrot model.  We will then discuss the interior  geometry of the
black non-inflating region in each phase which goes beyond the scope of
the model.  We regard this study as a first step towards understanding
various phases in the holographic dual theory, and also towards finding observational consequences of each phase.

\setcounter{equation}{0}
\section{Topological Order Parameters}
As explained by Garriga, Guth, and Vilenkin \cite{GGV}, eternal inflation requires an initial condition which we impose by choosing a space-like surface on which  the initial condition is white-vacuum-everywhere. We may choose the initial slice to either be compact or infinite. For definiteness we will consider homogeneous initial conditions  on  infinite spatially flat slices of de Sitter space (see however Section 8 and the appendices).

 We  assume that the basic decay process is nucleation of black bubbles in the white background. After a bubble nucleates the bubble boundary spreads out, quickly approaching  the speed of light.
The effect of a black bubble will be to block out a region of the white vacuum and prevent it from inflating, but the nucleation event will not have much effect on the remaining white region, which is out of causal contact with the nucleation event.

We will  not assume that \cdl \ bubbles are isolated. As shown by Guth and Weinberg~\cite{weinberg}, bubble collisions inevitably occur. What goes on inside the black regions can be very complicated. It depends on the properties of the inflaton potential as well as  the details of the bubble collision process. But because the white portion of space
 is outside of causal contact with the bubbles,  the white region is simpler to analyze than the black region. For that reason we will follow Winitzki~\cite{Winitzki} and define the various phases in terms of the topological properties of the percolating structures in the white region. Specifically, we will define objects called white crossing surfaces  (WXS's) and white crossing curves (WXC's). The order parameters distinguishing the various phases will be defined in terms of the existence or nonexistence of these white crossing manifolds.  Such a topological characterization was already familiar to Mandelbrot~\cite{Mandelbrot}.  Chayes, Chayes, Grannan and Swindle~\cite{Mandelbrot} characterized the phases of the Mandelbrot model in this way.

To define these terms, we begin by
choosing a space-like foliation of the space-time in the future of the initial slice.
Consider a particular leaf of the  foliation. It is possible that on that leaf there exists an unbounded two-dimensional  surface that never leaves the white vacuum. Such a surface consists of a single simply-connected component, and cannot be bounded by any finite sphere: if it exists it crosses the entire leaf. A white surface of this type is called a WXS.

A similar definition of white crossing curves can be given: a white crossing curve (on a space-like leaf) is a one-dimensional curve that crosses the entire space without ever leaving the white region. It is obvious that if there are white crossing surfaces on a leaf, there must also be white crossing curves. The converse is not true; the existence of a WXC does not imply the existence of a WXS.
It can easily be proved that if a WXS or WXC exists on a leaf, it will exist on any earlier leaf.

From now on, when we say that a WXS (WXC) exists, we will not be referring
 to any particular leaf.
A WXS will be  said to exist if it exists on every space-like leaf. This is important because the creation of new black regions can obstruct the white crossing manifolds. Thus their existence implies that during the entire future evolution, the new black regions do not block the existence of  WXS's (WXC's).

On the other hand, if  WXS's (WXC's) do not exist on a leaf, they cannot exist on a later leaf. To say that  WXS's do not exist means that they do not exist on arbitrarily late leaves of the space-like foliation. An important fact, proved in Appendix~B, is that the existence of  WXS's (WXC's) does not depend on the choice of foliation. In other words, if  WXS's (WXC's) exist with one choice of foliation, they exist for all foliations.

\subsection{The Phases}

The three phases of eternal inflation that we will discuss are defined in the following way.

\begin{itemize}
\item The black island phase is characterized by the existence of  WXS's. It follows that  WXC's also exist.

    \item The tubular phase has  WXC's, but it does not have WXS's.

    \item The white island phase has no WXS's or WXC's. But all leaves contain  white regions which  form isolated islands.

        \item The aborted phase in which no white survives: the entire space becomes black. This is the only phase that does not eternally inflate.

            \end{itemize}

Roughly speaking, one can envision the black island phase to be sparsely
populated by isolated black islands which do not obstruct the existence
of white crossing surfaces. In the tubular phase the black regions
percolate and form an infinite  tube-like network that does obstruct the
WXS's. However it does not obstruct the WXC's. The white island phase is
such that the black regions obstruct both WXS's and WXC's.

More precisely, when we consider compact global slicings, our definition
of the existence (nonexistence) of white crossing manifolds is that
the probability for their existence at future infinity is finite (zero).
For noncompact slicings, the probability is either one or
zero. See Appendix~A.

Evidence for the existence and properties of these phases follows from rigorous results about the Mandelbrot model obtained  by Chayes,
 Chayes,  Grannan, and  Swindle \cite{Mandelbrot}.

\section{Mandelbrot Percolation Model}

The Mandelbrot percolation model captures a lot of the physics of eternal inflation,  but before describing it let us review ordinary percolation theory.

\subsection{Simple Percolation in Three Dimensions}
Consider an infinite cubic lattice in three dimensions composed of unit cubes.
Start with the entire lattice colored white. Next,  go through the lattice and independently, with probability $P,$
``paint'' each cube black. With probability $(1-P) $  leave it white.

In the case of very small $P$   the  white region occupies most of the space, and the black region is very sparse. There are  two-dimensional  white crossing-surfaces which cross the entire infinite lattice without ever leaving the white set. Such crossing surfaces exist with probability $1.$ Obviously there are also one-dimensional white crossing-curves. In brief, for small $P$ the system is in the black island phase.

Consider the black set of points in the black island phase.
The black set is formed entirely of disconnected bounded  islands, each of which is completely surrounded by white. The probability that a connected black region crosses the infinite  lattice  is zero.
There are no black crossing-surfaces (BXS) or black crossing-curves (BXC). The probability that an island has size larger than $L$   decreases  exponentially with $L$
\be
{\rm \bf probability} (L) \sim e ^{-{L \over L_0}}.
\ee
The correlation length $L_0$ depends on $P.$ It is finite in the black island phase.

Letting $P$ increase, the correlation length grows. One finds that there is a critical point $P_c < 1/2,$ at which the black set percolates, and $L_0$ diverges. For $P_c< P< (1-P_c) $  there are no crossing-surfaces of either white or black, but there are crossing-curves of both colors. Furthermore there is a single infinite  connected set of black (same for white) that covers a finite fraction of the entire lattice.

The point $P_c$ is a critical point in the sense that the average size of a black island  tends to infinity as $P$ approaches $P_c$. At the critical point the behavior is scale invariant and correlation functions have power law behavior.

\begin{figure}\begin{center}\includegraphics{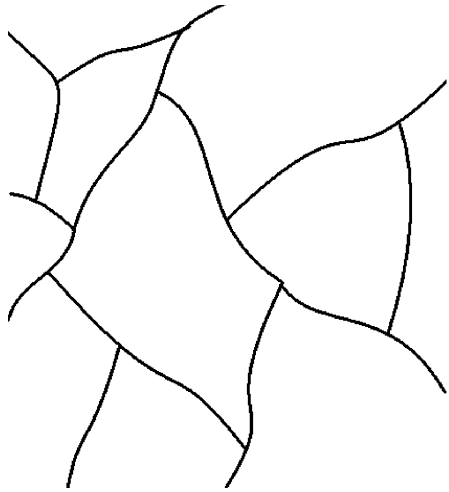}\caption{To construct a tubular  geometry begin with an infinite network of virtual strings.}\label{strings}
\end{center}\end{figure}

\begin{figure}\begin{center}\includegraphics{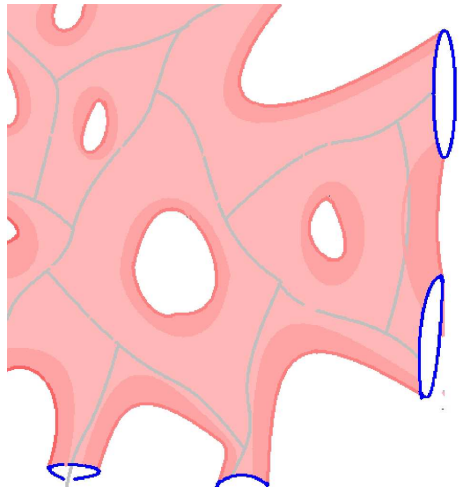}\caption{A tubular network built on a network of virtual strings.}\label{tubes}
\end{center}\end{figure}

The topology of the black  set in the tubular phase can be viewed in the following way. Begin with an infinite network of strings connected at junctions as in Figure \ref{strings}.
The string-network is  not composed of real strings. It is just a mathematical device to describe the black set. Once the string network is laid down, thicken the strings to form a  network of tubular regions as in Figure \ref{tubes}. The black set in this phase is the interior of the tubular network, plus a component that consists of islands surrounded by white. Note that there can be many islands, but there is only one infinite tubular set.

 The tubular set can be characterized by its boundary---a 2-dimensional orientable surface of infinite genus \cite{infinite riemann}.

This simple percolation problem  has an obvious symmetry
under the simultaneous exchange of $ {\rm white} \leftrightarrow {\rm
black}$ and $P \leftrightarrow (1-P)$. For example the theory 
at $P=1/2$ is symmetric under exchange of white and black. It follows
that the white set is also composed of an infinite network of tubes, 
and a collection of white islands within the black tubes. In fact the 
topological characteristics of the black and white sets are the same in 
the entire interval $P_c< P< (1-P_c) $.

For $P > 1-P_c$ the black set contains infinite black crossing surfaces (BXS), and the white set is composed of islands.

One can make various minor modifications of the rules---most modifications will break the  white-black symmetry---without changing the essential features of the model, namely
the existence of three phases: a black island phase, a tubular phase, and a white island phase, separated from each other by  second order transitions, as would be expected from the scaling behavior at the critical point.
A particularly simple modification is to begin with an empty white three-dimensional space. At each point we can lay down a black ball (disc in 2-D) of unit radius. The balls are laid down completely randomly with a probability density $\rho.$  When two balls overlap the overlapping region is colored black. The control parameter replacing $P$ is $\rho,$ and as it is varied from zero to infinity the system passes through the black island, tubular, and white island phases.

\subsection{Mandelbrot Percolation }

Again begin  with an infinite cubic lattice composed of unit cubes. Let all the cubes begin their existence in the white phase. As in simple percolation,   go through the lattice and with probability $P$  paint each cube black: with probability $(1-P)$  leave it white\footnote{Our notation is not the same as in \cite{Mandelbrot}, in which the notation $P$ is used for the probability that a cube remains white. To compare this paper with \cite{Mandelbrot} one should replace $P$ by $(1-P).$}. Once a cube is painted black, it stays that way forever. In a sense it dies and cannot reproduce.

In the next step we subdivide each white cube into $8$ smaller cubes. In other words we subdivide the white set into a lattice of half the original spacing\footnote{In \cite{Mandelbrot} there are a family of models parameterized by an integer $N$ in which the white cubes are subdivided into $N^3$ cubes. There is no important difference between the model with $N=2$ and models with larger $N.$}.
 Then we go through the new white array of cubes, coloring each one black with probability $P$.

This process is repeated ad infinitum\footnote{An alternate procedure is to double the size of the lattice at each step, so that the smallest cubes are always of unit size. This makes the connection with inflation manifest. }. Figure \ref{checkers} illustrates the first few steps (for the analogous 2-dimensional model). At each stage the  white cubes of linear size $2^{-n}$ are divided into  cubes of size $2^{-(n+1)}$which are then painted black with probability $P$. The black cubes remain black.

After $n$ iterations the white set of points is called $A_n.$ Thus at the beginning the white set, $A_0,$ comprises the entire space. After one iteration the  white set is $A_1,$ etc. The fraction of the original volume occupied by $A_1$ is $1-P$. After $n$ iterations the fraction of volume occupied by $A_n$ is $(1-P)^n.$

At the end of the entire process the white set has vanishing volume but  is not empty. It is a fractal with fractal dimension $d<3.$

\bigskip
\bigskip

The properties of Mandelbrot percolation include the following.

\begin{itemize}
\item For any finite $n$ the properties are the same as for simple percolation. There are two second order transitions at $P=P_1(n)$ and $P=P_2(n)$ separating the  white and black island phases from the intermediate tubular phase.

    \begin{figure}\begin{center}\includegraphics{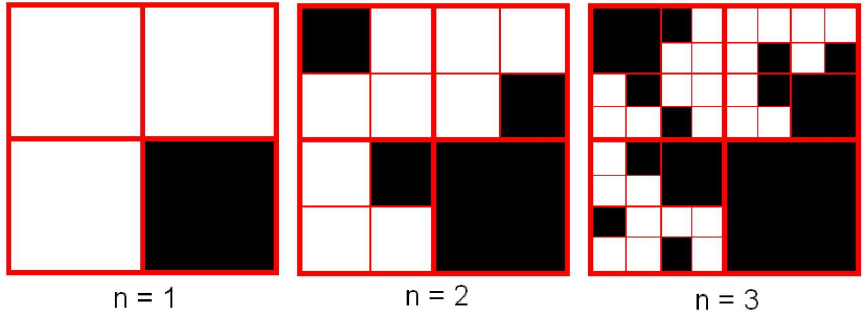}\caption{The first three iterations of the 2-dimensional Mandelbrot percolation model for $P = 1/4$.}\label{checkers}
\end{center}\end{figure}

    \item Both transition points are monotonically decreasing with $n$:
    \bea
    P_1(n+1) &\leq& P_1(n),  \cr
       P_2(n+1) &\leq& P_2(n).
       \label{inequality 1}
    \eea

 \item  The limits $P_i(\infty)$ exist:
\bea
 P_1(\infty)&>& 0, \cr
 P_1(\infty)&<&  P_2(\infty) < 1.
   \label{inequality 2}
 \eea

 The inequalities (\ref{inequality 1}) just come from the fact that in going from $n$ to $n+1$ new  black points may appear and none are removed.
 The inequalities (\ref{inequality 2}) are proved in \cite{Mandelbrot}.
 Thus the existence of three phases is inherited from  simple percolation.

 \item For $n=\infty,$ and for every value of $P,$ the white set, $A_{\infty},$ is a fractal at small distances with a fractal dimension less than 3.

      \item  The transitions at $n=\infty$ are first order. Unlike simple percolation, correlation functions do not become long range as the transition points $P_i(\infty)$ are approached.

          \item In the tubular phase there are no islands of black in the limit $n=\infty$. The black islands  which  exist for finite $n$ become connected  by smaller thread-like sets of points as $n \to \infty$. The black region becomes one big connected tubular structure.
     \end{itemize}

\subsection{ Cracking in the Mandelbrot Model}

\begin{figure}\begin{center}\includegraphics{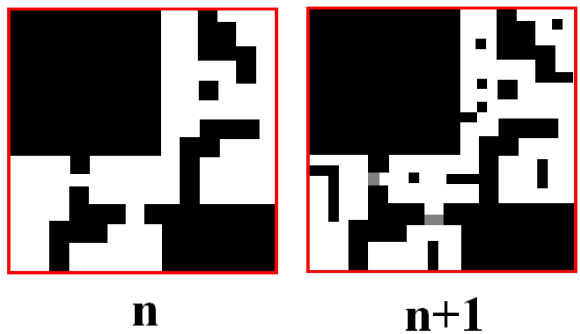}\caption{Stages $n$ and $n+1$ of the two-dimensional Mandelbrot model. The newly produced cells that cause the cracking are colored grey  instead of black for easy identification. }\label{crack}
\end{center}\end{figure}

In passing from the tubular to the white island phase a phenomenon called cracking occurs. Cracking occurs when going from stage $n$ to $n+1$ if a newly formed black region cuts through a white region thereby cutting it in  two. Cracking is illustrated in Figure \ref{crack}. The figure shows the  configuration at stages $n$ and $n+1,$ with a crack developing.

As $n$ increases cracking is an ongoing process that keeps splitting the inflating white islands into smaller and smaller islands. At stage $n$  the largest white islands have a size of order $2^{-n}.$  Thus, in  the asymptotic limit in the white island phase, the white set consists of a dust of points of vanishing size.

\bigskip

    As in the case of simple percolation, the Mandelbrot model can be modified without essential change in the behavior. As an example, stage one would be the same as described earlier: unit black balls (or discs) are randomly laid down on a white background with probability $\rho$. At the next stage, $n=2,$ black balls of radius $1/2$ are laid down with probability $8\rho$ ($4\rho$ in two-D). At stage $n$ balls of radius $2^{-n} $ are laid down with probability $8^n$. Regions where balls overlap are painted black. The phase structure is the same as the lattice Mandelbrot model.

\setcounter{equation}{0}
\section{The Black Island Phase}

Let us now return to eternal inflation, driven by the potential in Figure \ref{potential2}. The behaviors for very small and very large $X$ are familiar.
For small $X$ the cosmology consists of an unimpeded slide to the black vacuum: the universe is described by conventional FLRW cosmology with $k=0$ or $k=1.$ For the flat case the universe eternally expands at a decelerated rate; for the closed case it ends in a singularity after a finite time.

For very large $X$ the universe will be trapped in the white minimum,  inflating forever. Classically the geometry will approach de Sitter space,
 with a Hubble expansion rate $H,$
\be
H= \left({8\pi G\over 3}V_w \right)^{1/2}.
\ee
In the flat slicing the metric of \ds \ is
\be
ds^2={1\over H^2 T^2} ( -dT^2 +   dx^i dx^i)
\label{ds}
\ee
where $T$ is conformal time. The range of $T$ is $T<0$. The surface  $T =0$ will play a fundamental role in what follows. Note that it corresponds to the future infinite boundary of de Sitter space.

 As in ordinary quantum field theory,  \cdl \ tunneling transitions will
 occur with rate  $\Gamma$, producing bubbles of black vacuum.  This
 time however the transition will only reach completion if the
 nucleation rate $\Gamma$ is larger than the fourth power of the
 expansion  rate, $\Gamma \gtsim H^4$. If $\Gamma \ltsim H^4,$ the
 expanding bubbles will not outrun the hubble expansion and cannot
 coalesce. Inflation in the regions between the bubbles will be eternal.

 A dimensionless version of the nucleation rate is defined by
\be
\gamma = \Gamma H^{-4}.
\label{gamma}
\ee
and the condition for eternal inflation is
\be
\gamma  \ltsim 1.
\ee

\begin{figure}\begin{center}\includegraphics{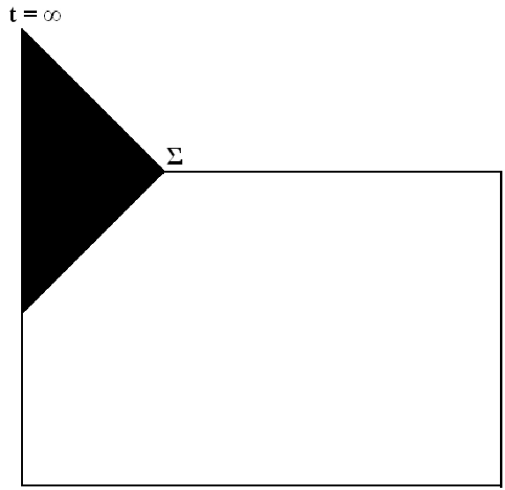}\caption{Penrose diagram for the nucleation of a bubble with vanishing \cc.}\label{penrose1}
\end{center}\end{figure}

Following Garriga, Guth, and Vilenkin \cite{GGV} we introduce an initial condition such  that at some (negative) value of $T$ all of space is filled with white vacuum. Without loss of generality we can choose the initial condition at $T= -1.$ As time evolves
 bubbles of black vacuum will nucleate.  As in the percolation model, we color the false vacuum white and the true vacuum black. Let us follow the history of a single  bubble that nucleates at $(T = -1, \ x=0).$ The process is shown in the Penrose diagram, Figure \ref{penrose1}.

\begin{figure}\begin{center}\includegraphics{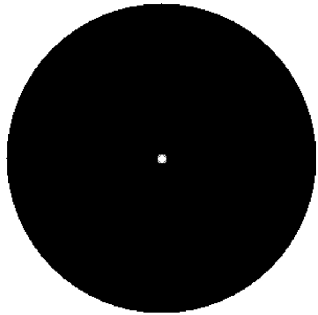}\caption{A single \cdl \ nucleation seen at time-like infinity.  The black region represents the light-like ``hat" in Figure \ref{penrose1}. The dot at the center of the black region represents the tip of the hat, i.e., $t=\infty.$}\label{bubble}
\end{center}\end{figure}

 The bubble is bounded by a domain wall which grows with time, eventually approaching the light-cone
 \be
 x^2 = (1+T)^2.
 \label{lightcone}
 \ee
 By the time  the bubble wall reaches $T=0$ it will be of infinite  proper size, but its comoving size will be $|x| = 1$. From the viewpoint of the $T=0$ surface the bubble cuts out a black comoving ball from the inflating white background as in Figure \ref{bubble}.
 The boundary of the black region at $T=0$ is called $\Sigma$.


The interior of the bubble is described as an open FLRW universe with metric
\be
ds^2 = -dt^2 + a(t)^2 d{\h3}^2.
\label{frw}
\ee
The space-like slices  $\h3$ are negatively curved three-dimensional hyperbolic planes
 that
are isomorphic to three-dimensional Euclidean anti de Sitter space.
\be
d\h3^2 = dR^2 + \sinh^2R \ (d\theta^2 + \sin^2\theta \ d\phi^2).
\label{ads}
\ee
At early and late times the scale factor behaves like
\bea
a(t) \eq t \ \ \ \ (t \to 0)  \cr
a(t) \eq t + c \ \ \ \ (t \to \infty)
\eea
where $c$ is a constant determined by integrating the FLRW equation of motion.
From within the bubble, the surface $\Sigma,$  has the meaning of  the asymptotic boundary of space at $R=\infty.$

Note that the later a bubble nucleates, the smaller its image will appear on the $T=0$ surface. For example, nucleation at time $T$ excises a  sphere of size $|T|$ from the white portion of the $T=0$ surface, and colors it black. It is the properties of the final pattern of black and white point-sets on the future boundary that define the phases of eternal inflation.

 Bubble  nucleation is a perpetual process that continues into the
 infinite future, so that the view from $T=0$ will be a self-similar
 fractal \cite{weinberg, AryalVilenkin}. More exactly, the white set will be a fractal of
 dimension less than 3. As we will see, it occupies zero measure of the
 comoving volume. However the \it{proper} \rm volume of the white set
 (on $T=0$) will be infinite.

As shown by Guth and Weinberg
  \cite{weinberg}, 
every bubble will eventually collide with an infinite number of smaller bubbles.
 In terms of the pattern of bubbles seen at $T= 0$, the black island phase for extremely  small nucleation rate
 looks like a collection of disconnected black islands. The largest size
 island consists of a bubble nucleated earliest and bubbles that
 collided with it. Smaller islands correspond to later nucleations.
Apart from an infrared cutoff due to the initial conditions, the distribution of bubbles will be scale invariant~\cite{BenMatt}.

 In the interior of the black bubble regions the geometry is entirely different than it would have been had the bubble not nucleated. But the inflating regions outside the bubble are not appreciably affected by the nucleation, especially if the  critical bubble size \cite{Coleman:1980aw} is much smaller than $H^{-1}.$ In that case almost all of the white set is causally disconnected from the nucleation and is therefore relatively easy to model. We will make the approximation that the metric is given by (\ref{ds}) everywhere outside the bubbles. The conformal time $T$ will only be used in the inflating region, i.e., not in bubbles.

\begin{figure}\begin{center}\includegraphics{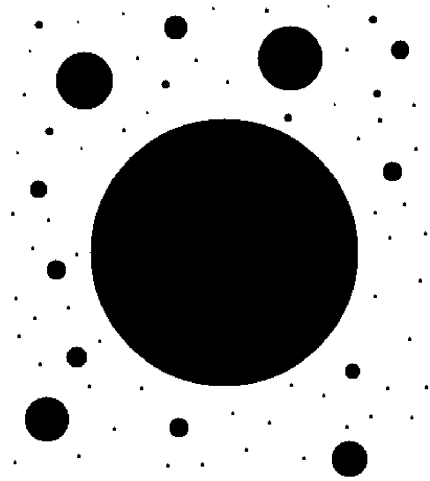}\caption{Bubble  nucleation is a perpetual process. The later a bubble nucleates the smaller it will appear on the $T=0$ surface.}\label{bubbly}
\end{center}\end{figure}

Typically the  number of bubbles in a comoving volume larger than $\Delta x = 1$ will be infinite, being dominated by arbitrarily small bubbles that nucleate arbitrarily late.
 This leads to infinities that we will need to regulate. To that end we introduce a regulator-time in the inflating  region. We define it by $T = -\e$. We will allow $\e$ to go to zero geometrically, i.e.,
 \be
\e =2^{-n}
\ee
 where $n$ is an integer. To remove the regulator we let $n\to\infty.$ The regulated distribution of bubbles consists of all bubbles that nucleate before the cutoff time, that is, $T<-\e.$ Examining the pattern of excised black regions on the future boundary, we find a finite number of bubbles in each comoving volume. See Figure \ref{bubbly}.

 Let us consider how much white comoving coordinate volume remains  at cutoff $\e$. Consider the fraction  that is eaten by black bubbles when the cutoff is $\e = \12.$ Call it $1-f$. The fraction $1-f$ is proportional to the nucleation rate $\gamma.$ The fraction  colored white is $f$. During each factor of $2$ decrease in the cutoff the fraction of white is decreased by fraction $f$. Thus for $\e =2^{-n}$ the remaining fraction of white is
 \be
 f_n = f^n \to 0.
 \label{fn}
 \ee
 The set of white comoving points that is left as $n\to \infty$ is a
 fractal \cite{weinberg, AryalVilenkin} of measure zero, similar to a Cantor set except that it is not defined deterministically but rather statistically.

\begin{figure}\begin{center}\includegraphics{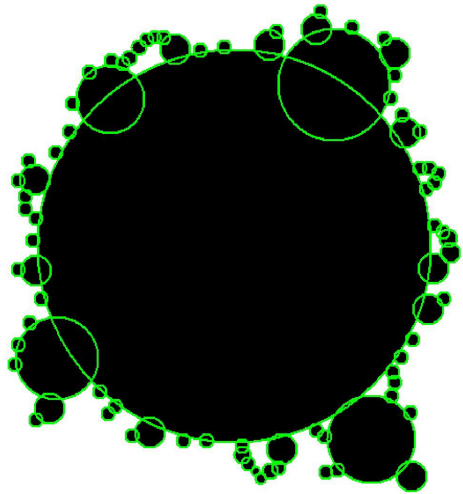}\caption{Bubble  collisions create multiply connected fractal  boundaries.}\label{froth}
\end{center}\end{figure}

 However this is not the whole story.  Bubble collisions have two important consequences, both easy to see in the Mandelbrot model. First of all bubble collisions can lead to white islands being trapped inside black regions. When this occurs the boundary of a black island becomes disconnected, although the island itself is connected. For the moment we will ignore such white islands and return to them in Section   7. The second effect is on the outer boundary, $\Sigma,$ of the black island. Although $\Sigma$ will remain connected, it will become multiply connected, and will develop a non-zero genus as illustrated in Figure \ref{froth}.

The similarity between Mandelbrot percolation and eternal inflation is clearest  if we focus on the $T=0$ surface. Bubbles nucleated between $T=-1$ and $T=-\12$ are analogous to the first generation of black cubes. They block out spheres  of coordinate size $\sim 1$ on the $T=0$ surface, in much the same way that the first generation black cubes do. Bubbles that nucleate between $T=-\12$ and $T=-{1 \over 4}$ have the same effect as the second generation of black cubes, and so on.
The probability $P,$ for coloring a cube black, plays the role of the nucleation rate $\gamma,$ and
the sequence of cutoff theories with $\e = 2^{-n},$ is the analog of the sequence of percolation models with finite $n$.

What the Mandelbrot model cannot  tell us is the behavior inside a black bubble. For that we need to study the geometry of such  bubbles.

\section{ Inside Black Islands }

Thus far we have characterized the various phases of eternal inflation by the properties of the eternally inflating region.  From a phenomenological viewpoint it is obviously more relevant to describe the phases from the viewpoint of the black region. To do this in any generality  is a more difficult task that we hope to  come back to. In this paper we will make some incomplete observations about the nature of each phase as seen from the black region.

Black islands are relatively simple. In the limit of small nucleation
rate they are small deformations of \cdl \ bubbles which are described as open FLRW universes. Consider a single \cdl \  tunneling event.  The intersection of such a bubble with the $T=0$ surface  is topologically a 2-sphere called  $\Sigma$ in \cite{FSSY,hat}. The cosmology within each bubble is a negatively curved  FLRW universe which becomes curvature dominated both at early and late times. The metric has the form
(\ref{frw}),
 where as we mentioned earlier, the space-like slices  $\h3$ are negatively curved hyperbolic planes,
 isomorphic to three-dimensional Euclidean anti de Sitter space.

The 2-sphere $\Sigma$ has several important properties. It is first of all the boundary separating the black and white sets on the future boundary. It is also space-like infinity from the viewpoint of the FLRW universe. As such, it is the natural candidate for the location of a holographic description of the observable universe within the bubble. A proposal for such a holographic theory was given in \cite{FSSY} and \cite{hat} in which the degrees of freedom  located on  $\Sigma$ include a Liouville sector and other ``matter" fields.

Another feature of $\Sigma$ is that it is the boundary of the ``hat." The hat is  light-like infinity---often called ${\cal{I}}^+$  or scri-plus---and the place where the generators of ${\cal{I}}^+$ enter the bulk geometry is $\Sigma$. Future time-like infinity is called the tip of the hat.

Bubble collisions define localized disturbances on $\Sigma$. The later the collision occurs, the smaller (in angular size) is the disturbed patch of $\Sigma$. In \cite{FSSY} we referred to these disturbances as 2-D instantons.

The simplest bubble collisions leave $\Sigma$ simply connected. But in general
bubble collisions  complicate the boundary and make it a higher genus surface
\cite{foam}.
For any Riemann surface of arbitrary genus, that surface can be the boundary of a three dimensional space with constant negative curvature. The construction can easily be understood. Let us start with a genus $1$ boundary---a 2-torus. The space $\h3$ can be described in terms of coordinates $z, r, \theta$ with
$$-\infty \leq z \leq \infty, \ \ \ \ \ 0 \leq r \leq \pi / 2, \ \ \ \ \ 0 \leq \theta \leq 2\pi.$$
\be
d\h3^2 = {1 \over \cos^2 r}(dz^2 + dr^2 +\sin^2 r \ d\theta^2).
\ee
The boundary $\Sigma$ is at $r=\pi/2$. In this presentation $\h3$ has the form of an infinite solid cylinder. To form the solid torus we do the obvious thing: allow $z$ to range over the finite interval from $z=0$ to $z=z_0$ (where $z_0 > 0$) and identify the discs at the two ends.
The identification can be done with a twist by an arbitrary angle $\alpha$  ($-\infty < \alpha < \infty$). Thus there are a two parameter family of such  tori, parameterized by the upper half plane $(z_0, \ \alpha).$

\begin{figure}\begin{center}\includegraphics{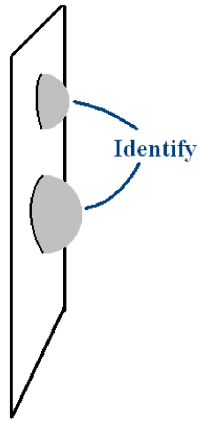}\caption{Construction of the solid torus geometry with genus $1$ boundary.}\label{ident}
\end{center}\end{figure}

Another way to do the same thing is to use Poincare coordinates in which the boundary is a plane at $U=0$,
 \be
ds^2 = {dZ dZ^{\ast} + dU^2 \over U^2 }  \ \ \ \ (U>0).
\ee
 We then cut out two solid hemispheres with their equators on the boundary, and identify as in Figure \ref{ident}.
 In this way we construct negatively curved solids bounded by 2-tori, $\Sigma.$ The 2-tori are parameterized by the Teichmuller space of genus $1$ surfaces. The usual modding-out  procedure that defines the moduli space of tori is not appropriate in this case. The reason is that a fat solid torus is not the same as a thin solid torus: nor is a twist by $2\pi$ the same as no twist.

More generally, $2g$ hemispherical holes can be introduced
in  pairs.  When the  holes are identified in pairs the result is a boundary of genus $g$.
We will call such geometries ${\h3}_g.$ For  $g>1$ the Teichmuller space of these geometries has real dimension $6g-6.$

For each such three-geometry we can construct a curvature dominated cosmology of
 the form
\be
ds^2 = -dt^2 + t^2 d{\h3}_g^2.
\label{topological cosmology}
\ee

The conjecture of \cite{foam} was that whenever a cluster of bubbles
(of the same vacuum) nucleates in such a way that the boundary
of the union of the bubbles has genus $g $ at the future conformal
infinity, the interior metric tends to (\ref{topological cosmology})
at late time. In \cite{foam}, explicit thin-wall solutions were found
which support this conjecture for the $g=1$ case, and heuristic argument
was given for general $g$.


\subsection{ Remarks About FRW/CFT}

Eventually we hope to describe the phases of eternal inflation in
holographic terms. Although we will not achieve that goal in this paper but
we find it worth discussing. Let us briefly review the ideas of
\cite{FSSY, hat}. The holographic formulation of an FLRW \cdl  \ bubble is in terms of a Liouville-matter two-dimensional field theory on $\Sigma$.  We begin with the simplest case in which   $\Sigma$ is topologically a two sphere with a geometry  determined by the Liouville field.

The CT's world-line is parameterized by a time variable which
we can take to be the usual FLRW time coordinate $t$. The CT
approaches the tip of the hat as $t\to \infty$.
At any given point on the world-line the CT looks back along his past light cone. As the CT's time increases that light cone moves out toward ${\cal{I}}^+.$

Now consider any 3-surface of fixed $t$. Apart from local
inhomogeneities the surface will be a uniform  hyperbolic space $\h3.$
The CT's past light-cone will intersect $\h3$ on a 2-sphere
$\Sigma_0$ which moves toward $\Sigma$ as the CT time increases.
This 2-sphere $\Sigma_0$ is the location of the regulated boundary
for the FRW/CFT correspondence; it plays the role of the ultraviolet
regulator in the boundary theory.
The evolution of CT time defines the RG flow of the boundary theory.

The holographic theory is a 2-D Euclidean field theory, which means that two coordinates, $R$ and $t$ must be emergent. These two dimensions are associated with the RG flow of the theory~\cite{hat}. As in ADS/CFT the coordinate $R$ is the reference renormalization scale. The time $t$  is identified with the Liouville  field and can be scanned by introducing a Lagrange multiplier to fix the Liouville field. The Lagrange multiplier is none other than the 2-D cosmological constant in the Liouville Lagrangian.

It is expected that as regulator goes to infinity---in other words as
the Census Taker time goes to infinity---the boundary theory decouples
into a pure Liouville theory and a matter CFT. The implication is that
at late time the conformal invariance of the matter theory becomes
exact. The conformal invariance of the boundary is the same as the
symmetry, $SO(3,1) $ of $\h3.$ In other words at late time perturbations
on the FLRW universe are diluted and the universe becomes homogeneous and isotropic on large scales.

At earlier times the theory is less symmetric. The RG flow starts at some initial bare action which is not conformally invariant and flows to a fixed point. This is a manifestation of the Persistence of Memory phenomenon of Garriga, Guth,  and Vilenkin \cite{GGV}. The implication is that at early time in the FLRW universe, translation invariance will be broken. Roughly speaking there will be a preferred center to $\h3$ and the CT should be able to observe it. At late time, as the RG flow tends to the fixed point, the asymmetry disappears.

Now let us consider more complicated topologies in which $\Sigma$
becomes multiply connected. Topological features can be described as
handles on $\Sigma.$ A statistical averaging over handles defines a
topological genus expansion similar to the genus expansion for the
world-sheet of string theory.
It is worth noting  that as  $n$ increases
the average genus of an island's boundary increases. One can think of 
the geometries as developing progressively smaller handles as in 
the Fischler-Susskind mechanism in string theory~\cite{fischler}.

In the black-island phase we expect that the topological disturbances,
i.e., handles, on $\Sigma$ will occupy small angular regions and the
later they occur the smaller they will be. We do not expect a
significant probability for handles which connect distant parts of
$\Sigma$. On the other hand the number of arbitrarily small handles will diverge. This is exactly the kind of divergence that occurs on the world-sheet of string theories and can be dealt with by renormalizing the local interactions on the world sheet \cite{fischler}. Such arbitrarily small handles do not threaten the locality of the boundary description. The tubular phase may be more problematic.

\section{The Tubular Phase}

As $X$ decreases and the bubble nucleation rate increases, a point will come where the black islands percolate. The effect of this  was demonstrated in the Mandelbrot model in which it was shown that for a range of $P$ the percolation of black clusters  obstructs white crossing surfaces, but it does not obstruct white crossing curves. In this phase, at finite $n$ both the white and black sets consist of  infinite tubular networks. In addition there are white and black islands trapped within the opposite-color tubes. However, as $n$ increases, the black islands become connected to the tubular network, so that the black set becomes a single connected infinite set.

In the black island phase the genus of an island's boundary only becomes
infinite when the cutoff parameter $n$ becomes infinite. By contrast, in
the tubular phase the genus of $\Sigma$  becomes  infinite  at finite $n$
(when we consider infinite slicings). The construction of geometries like ${\h3}_g$ can be extended to infinite genus, i.e., ${\h3}_{\infty}$ \cite{infinite riemann}. The obvious geometry for the black region would be
\be
ds^2 = -dt^2 +a(t)^2 d{\h3}_{\infty}^2.
\label{supertopological cosmology}
\ee

There is a very interesting point about multiply connected hats that was
 already described in \cite{foam}. In a multiply connected black region,
 one might have expected that any given observer (Census Taker)
 will only be able to see a portion of the connected component of $\Sigma, $ and of  its interior, no matter how long the Census Taker observes. However, this is not so. The causal past of any Census Taker  includes the entire multiply connected solid region ${\h3}_g.$ This  is especially true in the the tubular phase: any Census Taker eventually sees the entire (unique) multiply connected  FLRW geometry. The fact that the Causal patch of the Census Taker includes the entire boundary follows from two facts. The first is that in the simply connected case the Census Taker can eventually see all of ${\h3}.$ The second is that ${\h3}_g$ is obtained from ${\h3}$ by making identifications under a discrete subgroup of the symmetry group of ${\h3}$; namely, $O(3,1)$.

\begin{figure}\begin{center}\includegraphics{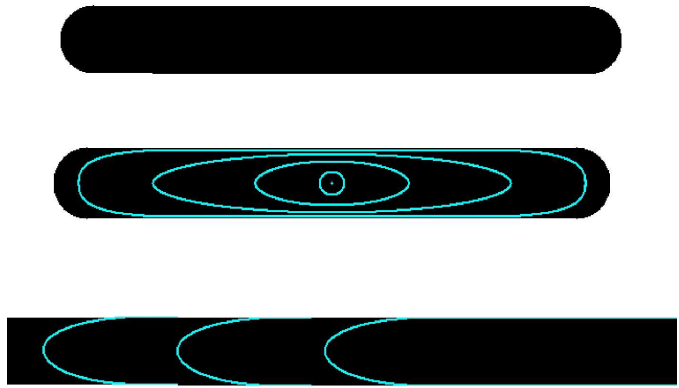}\caption{In the top picture, a long but finite black tube. The middle view shows the same black tube with the intersection of the CT's backward light-cones with an early equal time surface. The growing contours indicate how the intersection moves toward the boundary as the CT's time advances. The bottom picture shows a similar sequence for the case of an infinite tube.}\label{longtube}
\end{center}\end{figure}

 To describe what the CT sees in the various phases, and exactly how he
 would go about reconstructing the topology of the boundary is beyond
 the scope of this paper. 
However we can get some hints by temporarily ignoring the fractal nature of
the network and examining a couple of simple tube-like configurations.
First consider a long but finite tube created by the nucleation of a long, closely spaced, open chain of bubbles, nucleated along a line. On future infinity the bubbles will have coalesced and formed a long but finite solid cylinder with round ends as in the top picture in Figure \ref{longtube}. The boundary of the black region, $\Sigma,$  is a surface whose topology is a 2-sphere, but whose geometry is deformed into a long cylinder with round end-caps.

 Consider some particular space-like 3-dimensional slice $t=t_0$ in the tube.  We assume that the slice asymptotically tends to $\Sigma.$
Let us introduce a CT who moves along a particular time-like world-line. At any point along that world-line, the CT's backward light-cone will intersect the slice $t=t_0$ on a 2-dimensional surface $\Sigma_0$. For small Census Taker time, the intersection  forms a small 2-sphere centered near the CT's trajectory. As CT time advances, the sphere grows and becomes distorted and elongated. In the FRW/CFT formulation, the geometry of this surface is encoded in the Liouville field. A simple description of the geometry is a finite flat cylinder with curved end-caps. The integrated Gaussian curvature of each end-cap is $2\pi.$

As the CT time increases, the regulated surface $\Sigma_0$ moves toward $\Sigma.$ In this limit it is  expected  that the holographic field theory will become conformal and decouple from the Liouville  field. In other words at sufficiently late time, memory of the long tube-like distortion of the sphere will disappear. Let us briefly describe a simple  example of this decoupling.

Consider a 2-D massless scalar field $\phi$ in the  background geometry of a long 2-D tube. As is well known, a minimally coupled massless scalar Lagrangian does not couple to the Liouville field at all: it is invariant under Weyl re-scalings of the metric. Thus the minimally coupled scalar cannot distinguish the tube-like geometry from a spherical geometry.

But there is no general reason why the field $\phi$ should be coupled minimally. Let us suppose that $\phi$ couples to the 2-D gaussian curvature of the regulated boundary, through a term like
\be
\mu \int d^2x \sqrt{g^{(2)}}R^{(2)} V(\phi)
\label{curvcoup}
\ee
where $\mu$ is a numerical coupling constant, and $g^{(2)}$ and $R^{(2)}$ are the metric and gaussian curvature of $\Sigma_0.$ Now consider the  limit in which $\Sigma_0 \to \Sigma.$ In that limit the curvature $R^{(2)}$ tends to zero and the non-minimal coupling becomes unimportant.

But now consider  an infinite tube. In this case the surface $\Sigma_0$ does not uniformly approach $ \Sigma.$ From the third picture in Figure~\ref{longtube} one sees that the end-caps never  get close to the boundary. In  fact the curvature of the  end-caps remains constant. Thus in the model example, the curvature term in (\ref{curvcoup}) never becomes small.

To see the implications of this, we can map the infinite solid tube to the unit ball (3-D version of the Poincare  disc).  The asymptotic endpoints of the tube map to two points on the boundary. The two points can be chosen to be the north and south poles of the boundary sphere. The true gaussian curvature is concentrated on those points. The effect of that curvature is now reduced to a pair of insertions at the poles of the sphere. The insertions have the form
\be
2 \pi \mu  V(\phi)|_{\rm poles}
\ee

In terms of the RG flow of the Census Taker, the insertions tend toward local insertions as Census Taker time goes to infinity. At any given time the CT sees a pair of impurities on the sky, which shrink in subtended angle but never disappear.

Note that the impurities break the $O(3,1) $ symmetry of $\h3.$ One can  think of the  impurities as a  trail of debris running down the center of the  tube. The  debris is left over from the nucleation \cite{foam}. Our expectation is that such  trails  of debris will follow the tubular network in the tubular phase.
The fractal nature of the network will lead to a fractal pattern of
impurities on the CT's sky.

One point that should be noted is that the further away a signal originates, the more it is red-shifted by the time it is received by the CT.

To gain further insight into the tubular phase it  is interesting to
consider the solid torus. By studying it we get some idea of the effects
of multiple connectedness of the tubular network.
The solid torus can be thought of as a periodic infinite tube.  As the
CT time evolves, the same events on the surface $t=t_0$ will be repeated
over and over in both directions along the axis of the solid torus (see
Section 2.1 of \cite{foam}).
They will of course be seen as   progressively red shifted.

In practice observing the effects of living in a long tube, or of
multiple connectedness is severely limited. The comoving distance that
we can see to, on the surface of last scattering, is no bigger than $R
\sim .5,$ and then only if the number of e-foldings of conventional
inflation is minimal \cite{matt}. However, if we happened to be in an
unusually short fat torus,
it is possible that some effects might be visible.

There is another new aspect of the tubular phase.   As described above,
a finite CT time means a finite resolution for the CT's observations.
This serves as a cutoff, or lattice spacing, for the CFT.
At finite time the CT sees a regulated boundary, which is defined as
the intersection of the CT's past light cone and a spatial slice which
covers the tubular region. The genus of this regulated boundary
increases with time. At a given time there is a smallest handle
the CT can resolve.  As time goes on, the CT sees more and more
complicated boundary of the percolating network.

In the black island phase small handles typically reconnect to the dominant
island at distances of order the cutoff  and so give effects that can be
summarized by changes in local counterterms  in the
CFT~\cite{fischler}. But in the tubular phase a small tube at the cutoff
scale can connect to the big percolating network. It is not clear
whether the effect of such small handles in the tubular phase
can be summarized by a local CFT on Riemann surfaces.

 A somewhat analogous situation was studied in the context of ``wormhole" physics~\cite{coleman}.  There arbitrarily long thin wormholes connecting different points of spacetime produced an effect that was summarized not by a single local Lagrangian, but by a set of local Lagrangians with fluctuating coupling constants.  Here things are more complicated because of the fractal distribution of handle sizes.

\section{White Island Phase}

As the parameter $X$ in Figure \ref{potential3} decreases we should enter into the white island phase just before eternal inflation is aborted. It is very likely that this phase is closely related to slow-roll eternal inflation. The transition from no-eternal-inflation to slow-roll eternal inflation was described long ago by Linde and more recently by Creminelli, Dubovsky, Nicolis, Senatore and Zaldarriaga~\cite{leonardo}. Our expectation is that just beyond the transition the topology of the inflating region is island-like.

Mathematically the white  island phase is characterized by the absence of both white crossing surfaces and white crossing curves. Intuitively it is the phase in which the white set is composed of isolated white islands.
We have found this phase to be especially difficult to understand.  The
difficulties are partially associated with the continuous cracking (see
Section 3.3) that ultimately reduces the white set to a dust of
disconnected points. We will see that the cracking phenomenon is also
associated with the development of singularities on future infinity. We
suspect that these singularities cause observable black regions to
eventually undergo collapse similar to the big crunches of a positively
curved FLRW universe.

Our approach in this paper is to first turn off the bubble nucleation at
a finite time, let the geometry evolve into future infinity, then
discuss the effect of the cracking that happens after 
the cutoff time. Turning off the bubble nucleation is of course not a
consistent approximation. It is desirable to follow the real time 
evolution in a consistent manner, keeping track of the bubble nucleation 
at late times. Our conclusions about the white island phase should be 
taken to be tentative.

\subsection{Cracking Creates Singularities}

Before discussing the instability of  the white island phase, we will briefly discuss the stability of the black inland and tubular phases.  Consider these phases in the cutoff theory in which we terminate bubble nucleation after $n$ e-foldings in the inflating region. In the black island phase the topology of the boundaries $\Sigma$ will be finite and the geometry will evolve to the smooth non-singular geometry described in equation (\ref{topological cosmology}). As the cutoff increases, small handles develop and the genus increases, but at every stage there is a smooth candidate geometry to flow to.

The tubular phase is somewhat different. After some finite number of e-foldings, the black region will percolate. The boundary of the non-inflating  tubular network will remain connected but become of infinite genus. An infinite genus Riemann surface is not pathological and the geometry will tend to (\ref{supertopological cosmology}). Each such geometry is a negatively curved FLRW geometry and the curvature term in the FLRW equation should provide protection against collapse.

Beyond this point, increasing the cutoff will add smaller tubes to the network but at no point do we see any evidence of instability.

The situation in the white island phase is entirely different. This
phase is dominated by continuous crack formation, and as we will see, 
cracking is accompanied by gravitational collapse and the development 
of singularities.

We will approach the white island phase in stages. We will assume initial conditions on a global compact slice of de Sitter space. In the first stage we ignore all bubbles which nucleate after one e-folding (the case $n=1$).
 In the white island phase there is a significant probability of cracking in which the the globally white initial condition is split in two by a sheet or slab of black. This creates a boundary $\Sigma, $ which consists of two disconnected components.

In \cite{sasaki, foam}  a simple case was analyzed in which a crack creates two white islands  separated by a region of black. The boundary of the black region consists of two disconnected 2-spheres. One might think that an observer in the black region, located  between the boundaries, would eventually be able to see both of them. But this is not  the case.

What does happen is that
 the black region between the two disconnected boundary components is unstable, and collapses to form a  Kruskal-Schwarzschild geometry in which the boundaries are separated by a non-traversable \eb. Let us consider that example.

We begin with de Sitter space in the global slicing.  The metric is
\be
ds^2 = -d\tau^2 + {H^{-2}}\cosh^2 (H \tau) \ \ d\Omega_3^2
\ee
where $\Omega_3$ is the  unit 3-sphere,
\be
d\Omega^2 = d\alpha^2 + \cos^2 \alpha \ \ d\Omega_2^2.
\label{global ds}
\ee
In equation (\ref{global ds}) $\Omega_2$ is the unit 2-sphere and
$$-{\pi \over 2}< \alpha < {\pi \over 2}.$$

Consider the space-like two dimensional surface defined by the two conditions,
\bea
\tau \eq \tau_0, \cr
\alpha \eq 0.
\label{nucleation sphere}
\eea
Let us suppose that a collection of several bubble nucleations take place on or near that surface. Also assume that the nucleation events are more or less uniformly distributed on the surface, and that the spacing of nucleation events is less than the Hubble distance $H^{-1}.$
This, of course, is a very unlikely event but it is not forbidden.

Under these circumstances  the bubbles will coalesce and form a solid thick  sheet (or slab) of black vacuum. The black slab stays centered on $\alpha=0$ but grows thicker with time. The boundary of the slab has two disconnected components, each being a 2-sphere, which separate with the speed of light.   Eventually they reach future infinity at
$\alpha = a$ and $\alpha = -a.$

 Let us suppose that $a$ is just a little smaller than its maximum value, namely, $\pi/2.$ In that case most of future infinity is eaten by black, leaving over only  two small white caps,
$ a < \alpha \leq \pi/2$ and $ -a > \alpha \geq  -\pi/2.$ Naively each white cap is an island surrounded by black.

\begin{figure}\begin{center}\includegraphics{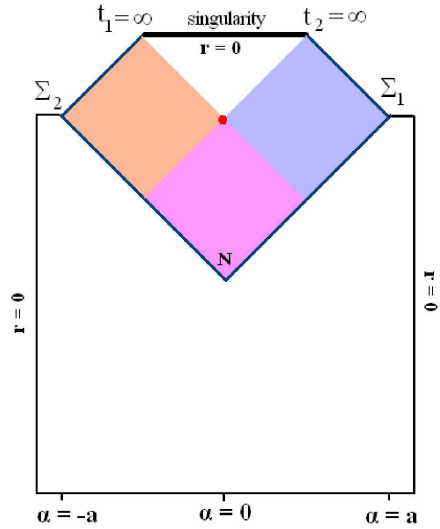}\caption{Penrose diagram for a surface-like nucleation event. $N$ represents the nucleation 2-sphere. The red dot represents the intersection of past and future horizons.}\label{schwarz}
\end{center}\end{figure}

There are a variety of pictorial ways to  visualize  the process, especially from the black side. The first is a Penrose diagram as in Figure \ref{schwarz}.
Following the diagram from bottom to top, we first see pure white de Sitter space. Then, on the 2-sphere labeled $N$ the sheet of bubbles nucleate and quickly thickens to form a slab between the two receding domain walls.  The boundaries eventually  reach $T=0$ at $\alpha = \pm a$. The asymptotic boundary $\Sigma $ now consists of two disconnected components, $\Sigma_1$ and $\Sigma_2$

 One can see the source of the instability by following a space-like  surface from $\Sigma_1$ to $\Sigma_2$. The boundary 2-spheres $\Sigma_{1,2}$ have infinite proper size. As we move away from  $\Sigma_1$ the local 2-sphere shrinks to a minimum and then expands to infinity as $\Sigma_2$ is approached. This sort of wormhole initial condition is guaranteed to collapse to a singularity.

 The problem of matching black and white regions across the domain wall
 was solved in \cite{sasaki, foam}. The solution is obtained by ``cutting and pasting" an ordinary Schwarzschild-Kruskal metric into the space  between the domain walls. In fact the upper portion of the diagram is nothing but the upper half of the full Kruskal diagram with two asymptotic regions connected by an \eb. The black region under  the hat has been split into two disconnected asymptotic regions, each with a black hole at its center.

 In \cite{foam} the mass of the black hole was determined by matching
 the solutions across the domain wall. In the limit of small tension,
 \be
 M = {R^3 \over 2 G l^2}
\label{BHmass}
 \ee
 where $R$ is the radius of the sphere defined in (\ref{nucleation
 sphere}) and $l$ is the radius of the de Sitter horizon.

There are two inequivalent Census Takers who asymptotically approach the
two time-like infinities $t_1 = \infty $ and $t_2 = \infty $. In other
words the two Census Takers reside on opposite sides of the crack and
each  can see only one boundary. The causal patches of the two time-like
infinities are shown in pale colors in Figure~\ref{schwarz}. The edges of the causal patches are the event horizons of the Census Takers.

The two caps between the poles and $\alpha = \pm a$ appear to be small and from the white viewpoint they are completely surrounded by black. However, from the black viewpoint
it is completely wrong to think of the black vacuum as surrounding localized white islands. Instead we have two disconnected black regions, each surrounded by a distant infinite sphere at $\Sigma_{1},\Sigma_{2}$ as in Figure \ref{schwarz}. 
There is a black hole at the center of a black region. The second
white region is behind the black hole horizon, just like the inflating universe
in an asymptotically flat space studied by Farhi and Guth~\cite{farhi}.

\begin{figure}\begin{center}\includegraphics{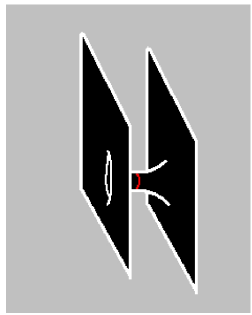}\caption{A two dimensional  analog of a  thick black slab apparently surrounding two white islands. The real geometry of the black set is more like an Einstein Rosen  bridge connecting two asymptotic regions. The red line represents the intersection of past and future horizons.}\label{bridge}
\end{center}\end{figure}

To help visualize the geometry on the  black side we can slice it from $\Sigma_1$
to $\Sigma_2, $ so as to pass through the intersection of the past and future horizons\footnote{Strictly speaking there is no past horizon. The red dot represents the place where the past and future horizon would intersect on the original Kruskal diagram.}, shown as a red dot in Figure~\ref{schwarz}. A two dimensional version of the resulting spatial geometry is shown in Figure~\ref{bridge}.

\begin{figure}\begin{center}\includegraphics{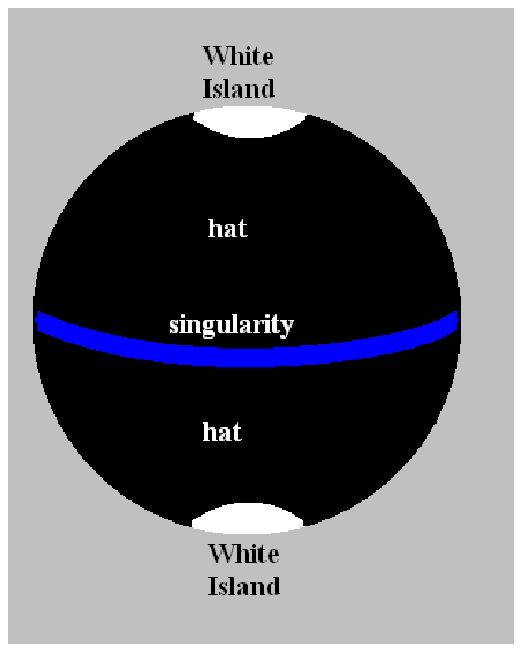}\caption{The asymptotic future of the geometry shown in the previous figure. The blue region represents the singularity that forms between two disconnected boundaries.} \label{sing}
\end{center}\end{figure}

 Another way to represent the  entire geometry is to focus on the future boundary. The future boundary  includes the surface $T=0$ in the  white  region, the light-like hats, and the space-like singularity. The topology of the future boundary is a 3-sphere. We of course color the white region white; the hat black; and the singularity blue. The result is shown in  Figure \ref{sing}. One could also replace the future boundary within the white region by a very late space-like surface. In that case the blue region would represent the portion of space inside the horizons of both types of CT's.

What a Census Taker sees on his sky will be the following: At first
he sees two domain walls (white islands) receding from him. One domain
wall (closer to him) moves away from him indefinitely. The other
moves towards the black hole horizon. From the CT's point of view,
the motion of the latter domain wall is red-shifted as it approaches
the horizon; eventually, it freezes on the horizon. The angular size
it occupies on the sky is finite. On the other hand, the former
domain wall continues to grow, and will eventually cover a large
portion of the CT's sky.

\begin{figure}\begin{center}\includegraphics{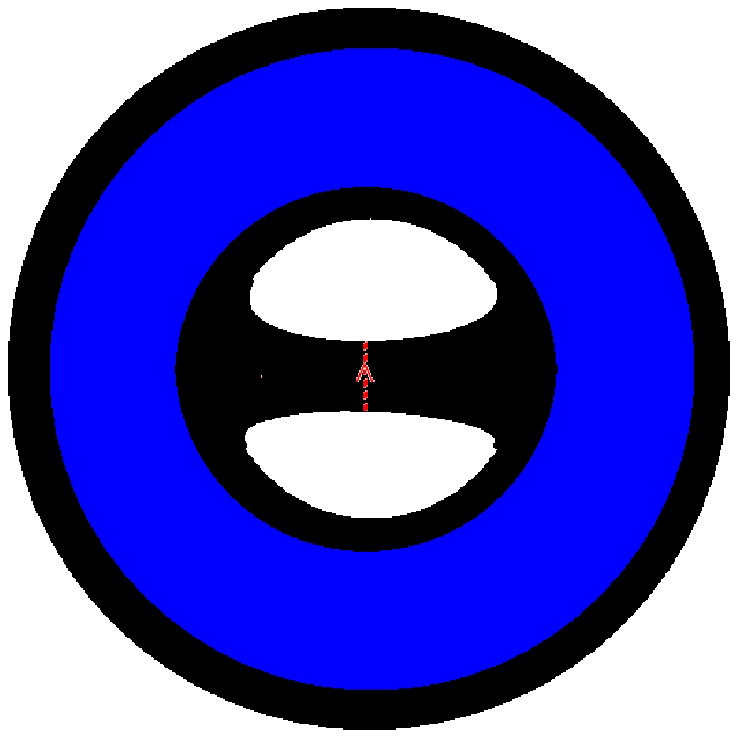}\caption{The white island cracks. The red dotted line represents a trajectory from one boundary to another.} \label{c2}
\end{center}\end{figure}

This of course is not the end of the story. In the next stage one or
both of the white islands may crack as in Figure \ref{c2}.

\begin{figure}\begin{center}\includegraphics{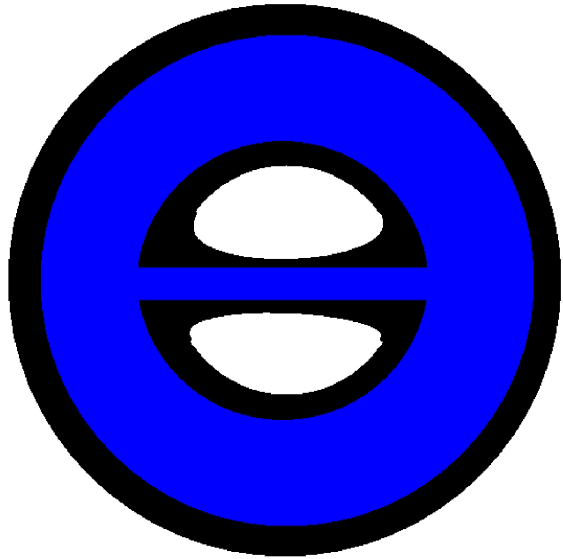}\caption{The absence of non-traversable wormholes implies that singularities must form that isolate the boundaries.} \label{c3}
\end{center}\end{figure}

\begin{figure}\begin{center}
\includegraphics{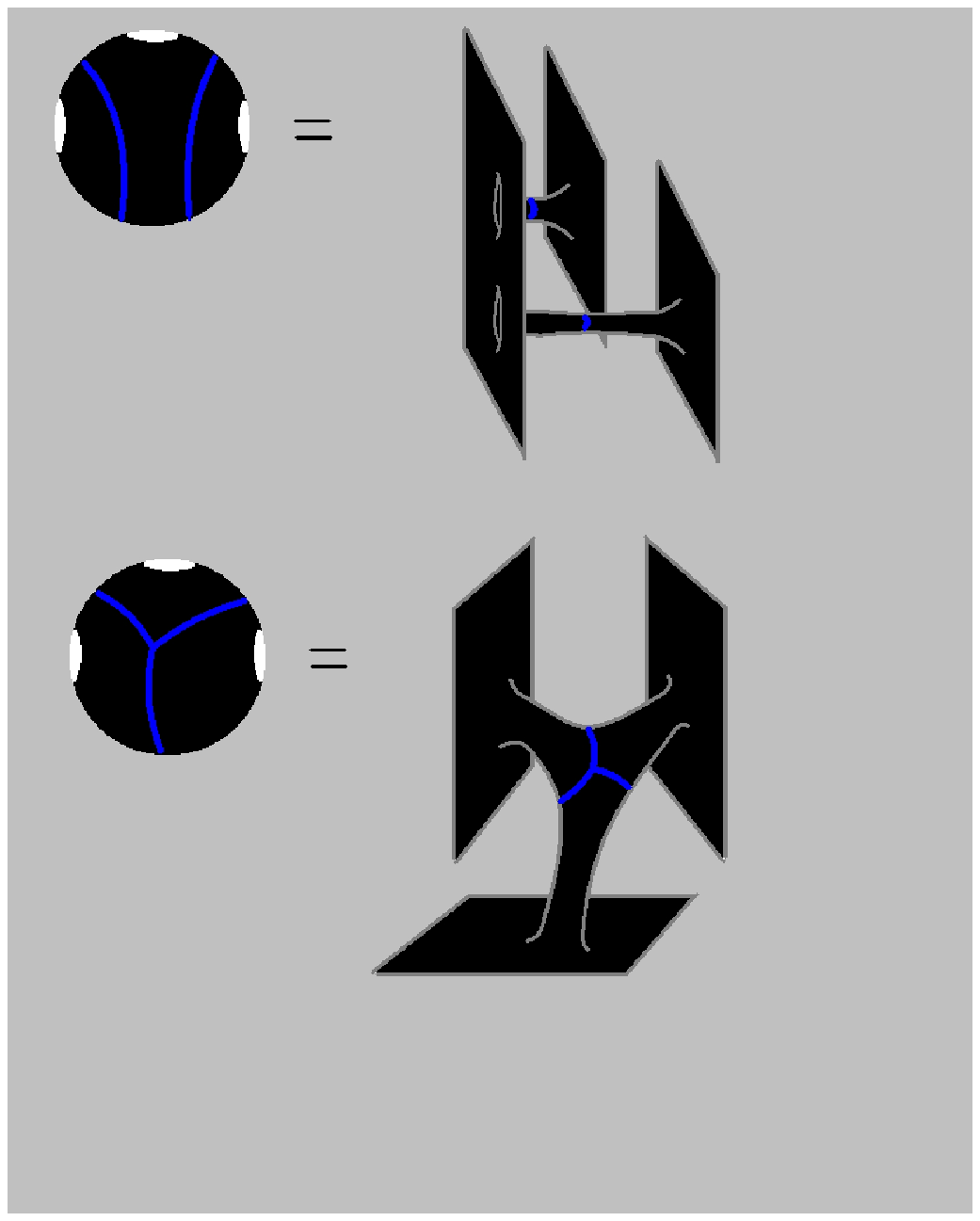}
\caption{A black
	       region surrounds three white islands. The blue region represents the singularity.} \label{triboro}
\end{center}\end{figure}

Recalling that the figures represent the asymptotic future, and that the metric in the inflating region diverges at late time, we can conclude that the metric on the two-dimensional boundaries (between white and black) diverges. Thus these boundaries represent asymptotic spatial infinity from the viewpoint of an observer in the black region.
The red dotted line in Figure \ref{c2} represents a trajectory from the vicinity of one boundary to another. If the geometry is regular along that trajectory there would be no obstacle to traversing the region between the two boundaries. One would conclude that there existed a traversable wormhole. Such wormholes can exist only if the null energy condition is violated. In this paper we will take a conservative view of traversable wormholes: we assume that they do not exist.

Thus one may conclude that any path connecting the two boundaries must
be interrupted by a singularity on the future boundary. One or the other
white regions must be surrounded by a singular (blue) region as in
Figures~\ref{c3}, \ref{triboro}.

This description can be extended to any fixed cutoff $n$ as long as we
shut off the nucleation beyond the cutoff time. The consequences of this
fact is that any Census Taker can asymptotically see only one connected
component of the boundary. He also sees at least one black hole. The
horizon of the black hole, when viewed from the asymptotic future, is
the  edge of the blue singularity. All the other asymptotic regions are
hidden behind the black hole horizons (see Figure~\ref{triboro}).

Let us consider the fate of an observer in the black region. One possibility is that the observer finds himself behind a horizon and winds up destroyed at a singularity. In that case we would not call the observer a Census Taker.
On the other hand the observer may remain in one of the black regions and approach time-like infinity, remaining outside the black  holes.

\subsection{Continued Cracking}
Thus far we have considered the evolution of the white island phase
assuming that bubble nucleation and  cracking cease at some cutoff
time. The picture we described involves following the system to the
infinite future after the cutoff time. However, this procedure is not
consistent.  As we have explained earlier, continued cracking causes the
white set to degenerate into a dust of points, each no larger than a few
times the cutoff length.    The meaning of this is that no white island
ever gets bigger than some finite number of Hubble lengths before it is
cracked. In other words, from the point of view  of any surviving Census
Taker, the boundary $\Sigma$ is always of order the Ancestor 
Hubble size.

The intuition from the spherically symmetric model suggests the
following picture.
At a given time, an observer in the black (non-inflating) region may see
more than one connected components of the boundary, each of order Hubble
size. But these components will go out of causal contact with each other
in finite time; they will be separated by a network of singularities and
horizons. By that time one connected component of the boundary may have
split into several pieces, and the above process will repeat. In a
(causally) connected piece of black region 
we will
find a finite number of Hubble-size boundaries and one or more black
hole horizons. In fact, because the horizon of a
black hole with mass (\ref{BHmass}) is of order Hubble scale 
and the size of the boundary is of the same order, we think 
it is inevitable that the horizons merge.

One possible interpretation is 
that the geometry within a connected black region never grows
large. Most likely it would collapse so that the entire black region is
replaced by a singular crunch. But it is possible that ordinary
inflation in the black region might cause it to inflate to a large size
before it collapses. The only way that we know  to reconcile a small
boundary with a large bulk geometry is to picture the universe as a
large three-sphere with a tiny hole in it, similar to an inflated
balloon which also has a  small boundary. In that case the observed
universe would resemble an FLRW universe with positive curvature. This
is an important issue to get straight since it would give an example of
something we have not seen before; namely, a mechanism for eternal
inflation to give rise to a universe with positive spatial
curvature\footnote{Andrei Linde has repeatedly emphasized the
possibility of the creation of closed universes in eternal inflation.}.

\subsection{Inflation Aborted}

The last phase transition---from white island to the aborted phase---is best defined by considering the initial conditions to be on a compact global slice of de Sitter phase. To model that situation we can consider Mandelbrot percolation on a finite space.
 As an example, start with a  single white  unit  cube, and then  divide it into eight cubes, color them black with probability $P,$ etc. For any value of $P$ the finite Mandelbrot model can abort at any finite stage if  every cube becomes black. For small $P$ the probability that it aborts as $n\to \infty$ is less than unity, but when   $P$ becomes too large the system will die (become all black) with probability $1$. Let us determine where the critical value of $P$ is at which eternal inflation is aborted.

After the first iteration the number of cubes is $8$ and the mean number of white cubes is $8(1-P)$. After $n$ iterations the number is $[8(1-P)]^n$. Thus the system will certainly die if $8(1-P) < 1.$ The critical value for aborting the process is
\be
P_a = {7 \over 8}.
\ee
For $P< P_a$ the probability for aborting is less than one.

Unlike the other transitions we have described, the transition from
white island to aborted is smooth. To define what we mean by smooth,
consider the probability that after an infinite number of steps the
white set has not completely disappeared. Call that probability
$\Pi(P)$. If $8(1-P) > 1,$ then $\Pi(P) >0,$ while for $8(1-P) < 1,$
$\Pi(P) =0.$ The question is whether $\Pi(P)$ is continuous or abruptly
jumps from zero to some finite value? It is easy to see
that the transition is smooth~\cite{branching}.

This model has obvious  application to the case in which the (white)  initial condition is specified on a compact global slice of de Sitter space.
If the parameter $X$ in Figure \ref{potential3}  becomes too small (the
decay rate too large) the transition to black will be completed, and
instead of eternally inflating the  universe will evolve  as a closed
FLRW universe.

\subsection{Connection With Slow-Roll Eternal Inflation}

\begin{figure}\begin{center}\includegraphics{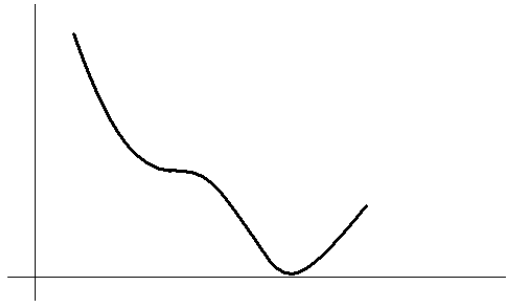}\caption{A potential for slow-roll eternal inflation.} \label{slow}
\end{center}\end{figure}

An alternative to bubble nucleation, that can also generate eternal inflation, is called slow-roll eternal inflation. As an example consider the potential in Figure \ref{slow}.  If the field starts on the plateau (analogous to white) then the rate at which thermalized (black) regions form will be of order $H$. It seems likely that the inflating white regions will be sparse and that the system, if it eternally inflates at all, will be in the white island phase.

 It is a very interesting question whether the view from within the thermalized regions is the same as we have described earlier.  In this regard we note that Bousso, Freivogel and Yang \cite{inside} have argued that slow-roll eternal inflation always leads to a crunch.

\section{Multiple Vacua}

The considerations of this paper are relevant in situations in which
only two vacua play an important role. When more than two vacua are important things become more complicated. As an example, consider the case of three vacua. We will call them white, grey, and black. White and grey have positive \cc \ and inflate; black has no vacuum energy. Just for illustrative purposes assume that  white can decay to grey with rate $\gamma_{wg}<<1,$ and grey can decay to black with rate $\gamma_{gb} \sim 1.$ Let us assume that the rate $\gamma_{gb} $ is such that decays from grey to black will produce a tubular grey-black phase.

There are many alternative descriptions of the phases of this system, but to keep the discussion close to the two-vacuum case, we will group white and grey together and define a white-grey set on the future boundary. The non-inflating black vacuum will define the black set.

Now consider what would happen if the universe started  white. It is evident that the black set must lie within the grey bubbles since only grey decays to black. Moreover, if $\gamma_{wg}$ is very small,  there will be WXS's. Thus the universe is in the  black island phase.

On the other hand let us focus on the interior of a grey bubble. Since the value of $\gamma_{gb}$ has been chosen to lie in the tubular range of the grey-black system, it is clear that the behavior within the bubble can be described as a grey-black tubular phase. For example there will be grey crossing curves (GXC) that cross the bubble.

We can model this situation with a Mandelbrot type model in an obvious way. At any stage each cube can be white, grey, or black. White and grey cubes reproduce by splitting into eight cubes. When a white cube splits its descendants are painted grey with probability $P_{wg}$ or white with probability $1-P_{wg}$. When a grey cube splits its descendants are painted black with probability $P_{gb}$ or grey with probability $1-P_{gb}$. As usual black remains black and does not reproduce.

\begin{figure}\begin{center}\includegraphics{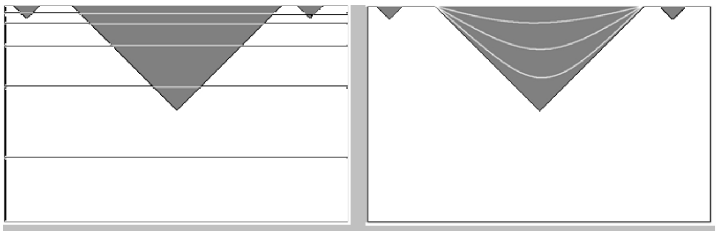}\caption{On the
	       left side the global geometry is foliated. The foliation
	       on the right covers only the grey bubble with FLRW equal
	       time slices. It does not cover the global geometry.
 } \label{folios}
\end{center}\end{figure}

To gain more insight into the multiple vacuum case, let us consider the interior of a grey bubble, momentarily ignoring the decay to black.
 A grey \cdl \ bubble can be described by an open, infinite, FLRW universe with a positive cosmological constant. The  homogeneous spatial hypersurfaces are the usual infinite hyperbolic planes $\h3.$ Asymptotically, the space inflates so that the FLRW scale factor increases exponentially. In Figure \ref{folios} the white vacuum is shown together with a number of grey bubbles.
 On the left the geometry is foliated by leaves that cover the global geometry, including the white and grey regions. When decays of grey to black are allowed, the black regions will be island-like since they are embedded in the grey bubbles.

 However, we get a different answer if we focus on a single grey bubble, and instead of foliating the global space, we foliate the  bubble  with surfaces of constant FLRW time.
 Such surfaces are infinite\footnote{ They are also typically hyperbolic. Simple percolation on a hyperbolic space is somewhat different than in flat space. But the differences should not be important to the asymptotic behavior governed by late  nucleating bubbles.}.
Since we have chosen $\gamma_{gb}$ to be in the tubular region, our definition would indicate that the bubble is in the tubular phase. This ambiguity seems to violate a claim made earlier and proved in Appendix B; namely, that the phase does not depend on the foliation.

 However, the two families of foliations cannot be compared in the manner required in Appendix B: they do not foliate the same regions, and can't be deformed into one another.
What is true  is that all global foliations will give the same phase---in this case black island---and all foliations of the FLRW bubble will give the same phase---tubular in this case.

\bigskip
\bigskip

This ambiguity in how the phase is described is a consequence of trying to characterize the system in terms of the properties of the inflating set.
It is clear that a Census Taker  in the black region has no such ambiguity. The CT looks into his past and as time unfolds, he will discover a tubular black environment.

This raises a question that we hope to come back to: Is it possible to give a  precise characterization of the various phases in terms that only refer to the properties of the black region?

\section*{Acknowledgments}
We are grateful to Ben Freivogel for crucial discussions, especially 
about the singularity formation.
We also thank Daniel Harlow, Jonathan Maltz and Vitaly Vanchurin
for very helpful discussions. The research of SS and LS is supported
by NSF Grant no.~0756174.
The research of YS is supported by MEXT
Grant-in-Aid for Young Scientists (B) No.~21740216.

\appendix

\section*{Appendix}

\section{Order parameter in the finite and infinite Mandelbrot model}

The two versions of the  Mandelbrot model---finite and infinite lattice---are appropriate for the global and infinite slicing of \ds.  The question we will answer in this appendix is, what is the relation between the infinite and finite models?

Ordinarily, statistical mechanical  models on a finite lattice  do not have phase  transitions. The different phases are only well defined in the  infinite limit. This is true of the simple percolation model. The Mandelbrot model is different in this respect. Not  only are there phase  transitions for the finite model, but they occur at exactly the same value of $P$ as in the infinite case.

For both the finite and the  infinite  model the phases are characterized by the existence of WXS's and WXC's. For the finite  model such surfaces are defined by crossing from one side of the system to the  opposite  side while remaining in the white set.

Let us consider the existence or non-existence of  a white crossing
surface (The arguments for WXC's are identical.).
The definition of the existence of such a surface does not refer to a particular value of $n.$
By definition a WXS exists iff  one exists for all $n.$

For the infinite model, the question of whether a WXS exists at a given value of $P$ is either yes or no. In other words the probability for their  existence is either zero or one, with the transition being at some critical $P.$ But in the finite model the situation is a little different: the existence of white crossings is a probabilistic matter. The reason is simple: in the finite model, no matter what the value of $P,$ there is a probability that at a given value of $n,$  an obstruction to the WXS  forms. The correct question is this: at a given value of $P,$ is the probability for a WXS zero or finite?
\begin{itemize}

\item Definition

$Q(P)$ is the probability that  no obstruction to WXS's  forms in the finite model. $Q(P)$ is monotonically decreasing with $P.$ It can be proved that at some non-zero $P,$ $Q(P)$ discontinuously jumps to zero. Call that value $P^{\ast}.$

\item Theorem

$P^{\ast}$ is identical to the point at which WXS cease to exist in the infinite model.

To prove the theorem, begin by defining a sequence of finite Mandelbrot models, $M_a,$ which begin with a lattice of size $2a \times   2a  \times 2a.$ The original finite model is $M_1,$ and the infinite model is the limit of $M_a,$ as $a\to \infty.$ We also define  $Q_a(P) $ to be the probability for a WXS in $M_a.$

 Consider a history of the finite model in which no black cubes are
      formed during the first $a$ steps. There is a finite probability
      $q_a$ for this to occur. From here on the problem is the same as $M_a.$
Obviously if $Q_{a}(P)\neq  0$ then $Q(P)$ cannot be zero, since
$Q_{a}(P)q_{a}\leq Q(P)$. Thus $Q(P)=0$
implies $Q_a(P) = 0$ for all $a.$
Thus if $Q(P)=0$ there are no WXS's in the infinite model.

Next suppose the
 probability for a crossing surface is non-zero in the finite model. Again consider a history in which the first n steps leave the lattice white. The probability that the subsequent history leaves a WXS is obviously greater than in the original problem on the $2\times   2  \times 2$ lattice. Therefore it follows that if $Q(P)>0,$ the probability for WXS in the infinite model is also greater than zero. But in the infinite model the probability is either zero or one.
Therefore it follows that the transition from the existence of WXS's to the nonexistence occurs at  the same value of $P$ in the finite and infinite models.

\end{itemize}

Exactly the same argument holds for white crossing curves. That proves that the phase structure of the finite and infinite models are the same.

\section{Slice independence of the phases}

When considering the definitions of the various phases of eternal inflation, we made reference to a foliation of the space-time by space-like surfaces.  This might be thought of as a form of gauge fixing. The question that arises is whether the distinction between phases is gauge invariant.
To put it another way, suppose a white crossing surface or curve exists
when a particular foliation of space-time is considered. Does it follow
that a WXS (WXC) also exists for any other foliation? The answer, explained below, is yes: their existence  is independent of the specific foliation. In what follows we will refer to WXS's but the arguments apply in exactly the same form to WXC's.

For the sake of definiteness we will restrict our discussion to the case of compact global slicing.
We will begin with two intuitively obvious lemmas.

\begin{itemize}
\item Definition

Let $L_1$ and $L_2$ be any two global space-like surfaces such that $L_2$ is everywhere in the future of $L_1.$ We indicate this condition by
\be
L_2 > L_1.
\ee

\item Lemma 1

If $L_2 > L_1$ and if $\exists$ a WXS on $L_2,$ then $\exists$ a WXS on $L_1.$

To see why this is true construct  a continuous family of interpolating surfaces between $L_1$ $L_2$ and consider how the topology of the black set varies as we scan the surfaces, proceeding from the later to the earlier surface. There are only two kinds of topological transitions that can take place. The first is that a simply connected island can shrink and disappear. This is the time reversal of a bubble nucleation.

The second is that a thin neck can pinch off. This is the time reverse of a bubble collision.

The point is that neither type of transition---disappearing or pinching off---can present a new obstruction to the existence of a WXS. Thus it follows that the existence of a WXS on $L_2$ implies the existence of a WXS on $L_1.$

\item Lemma 2

Consider any two space-like foliations of space-time  future to the
      initial condition.
 Call them $A$
and $B.$ Let the leaves of the foliations be denoted $L_a$ and $L_b.$  Then one can prove that
\be
\forall L_a \exists L_b : L_b > L_a
\ee
In other words for any leaf of $A,$ there exists a leaf of $B$ which is in the future of $L_a.$

This Lemma is only true for compact global slices of \ds.

\item Theorem

If a WXS exists (does not exist) for foliation $A$ then a WXS exists (does not exist) for foliation $B.$

Suppose a WXS exists for foliation $A.$ This means that for all leaves of $A$ there are WXS's. Now take any $L_b.$ From lemma 2 there is an $L_a$ in the future of $L_b$. Since it has a WXS lemma 1 implies that $L_b$ also has a WXS. Thus all $L_b$ have WXS's.

Next suppose that WXS do not exist for  foliation $A.$ This means that late enough $L_a$ do not have WXS's. Pick such an $L_a$ which does not have a WXS. Then by lemma 1, any $L_b > L_a$ cannot have a WXS.
This proves that the existence of WXS's is not dependent on the choice of foliation.

Since the defining properties of the black island, tubular, and white island, phases are in terms of the existence of white crossing surfaces and curves, it follows that these phases are gauge invariant under changes of foliation.

\end{itemize}

\end{document}